\documentstyle[preprint,aps,prl,epsf]{revtex}

\begin{document}

%\vspace{-2cm}
\title{Effects of Pore Walls and Randomness on Phase Transitions
in Porous Media}

\author{Marek Cieplak$^{1,2}$, Amos Maritan$^{3}$, 
Michael R. Swift$^{4}$, Flavio Toigo$^{5}$,
and\\ Jayanth R. Banavar$^{1}$}
\address{$^1$ Department of Physics and Materials Research Institute,
104 Davey Laboratory, The Pennsylvania State University, University
Park, Pennsylvania 16802}
\address{$^2$Institute of Physics, Polish Academy of Sciences,
Aleja Lotnik\'ow 32/46, 02-668 Warsaw, Poland}
\address{$^3$ International School for Advanced Studies (SISSA),
Via Beirut 2-4, 34014 Trieste, INFM and the Abdus Salam
International Center for Theoretical Physics, Trieste, Italy}
\address{$^{4}$ School of Physics and Astronomy, University of
Nottingham, Nottingham, NG7 2RD, UK}
\address{$^{5}$ Dipartimento di Fisica, Universita di Padova, Padova 35131, 
Italy, and Istituto Nazionale di Fisica della Materia, Italy } 
\address{
\centering{
\medskip\em
{}~\\
\begin{minipage}{14cm}
We study spin models within the mean field approximation
to elucidate the topology of the phase diagrams of systems 
modeling the liquid-vapor transition and the separation
of He$^3$--He$^4$ mixtures in periodic porous media.
These topologies are found to be identical to those of
the corresponding random field and random
anisotropy spin systems with a bimodal distribution of the
randomness. Our results suggest that the presence of walls 
(periodic or otherwise) are a key factor determining
the nature of the phase diagram in porous media.
%%%A simplified model for $^3$He-$^4$He mixtures in aerogel
%%%reproduces many of the features
%%%observed in experiments and suggests the opening of a
%%%miscibility gap at low $^4$He concentration.
{}~\\
{}~\\
%{\noindent PACS numbers:  71.28.+d, 71.27.+a}
\end{minipage}
}}

\maketitle

\newpage

\section{Introduction}
%\vspace*{-2cm}

Critical phenomena are generally well understood but the effects of randomness
on the nature of the transitions are less well-studied.
This is especially true in the case of phase transitions
that take place in porous media where the effects of quenched 
randomness are provided by the pore walls.
Among the best studied, are phase transitions in highly
porous aerogels \cite{Moses} ---
both the liquid - vapor transition \cite{Moses1}
and the $\lambda$ - transition of $^4$He \cite{Moses,Reppy,Reppy1}
have been found to be remarkably sharp. Even
more interestingly, the topology of the phase
diagram for $^3$He-$^4$He mixtures in aerogels has been found to be
different from that in the bulk \cite{Moses,Ma}.\\

The simplest theoretical framework for
studies of critical phenomena in non-random systems
is the Ising model. The important role played by multiple
length scales at a critical temperature $T_c$ leads to
universality \cite{Kadanoff,Wilson,Stanley} ---
binary alloys which are about to order, binary liquids which are
about to phase separate, certain kinds of magnets with uniaxial
anisotropy which are about to become magnetized all exhibit
the same critical behavior as the Ising model.
Perhaps the simplest extension that incorporates 
randomness is the random field Ising model
(RFIM) wherein a quenched random field is applied at each 
site \cite{Imry,Khurana,Young}. 
One example of the probability distribution is the symmetric bimodal
distribution which corresponds to a situation in which
half of the sites experience an up field and the other
half, a down field of equal strength.
Another example involves fields which are 
Gaussian-distributed. In both cases, the symmetry between the up and
down directions is not broken by hand and thus provides a scope for a
spontaneous symmetry breaking and a phase transition associated with it.\\

Recent research \cite{Swift,Sourlas,Usadel,Nowak,Duxbury} has lead to
the result that the two probability distributions may correspond,
at least in dimensions 3 and larger,
to distinct behaviors associated with the RFIM. The 
Gaussian case is governed by a $T$=0 fixed point while the bimodal
model's phase diagram is qualitatively different.
The origins of the two distinct scenarios relate to their
quite different $T$=0 phase diagrams.\\

Experimental realizations of the RFIM 
include dilute
antiferromagnets in a uniform field \cite{Birgeneau,Wong,Belanger,Hill}
and binary liquid mixtures in a porous medium
\cite{Maher,Goh,Dierker,Wiltzius,Wiltzius1,Cannell,Aliev,Sinha}. 
In both these cases, many of the expected signatures
associated with the RFIM with a Gaussian distribution of random fields
were observed \cite{Birgeneau,Wong,Belanger,Hill} (but significant
deviations were also found when the disorder was 
correlated \cite{Hill}).
However, the sluggish dynamics
and irreversibility predicted by the theory \cite{Khurana,Young,Fisher1}
precluded accurate
measurements of the exponents for binary liquids in porous media.
Nevertheless, the exponents determined for the dilute Ising antiferromagnet
in a uniform field were in accord with the theory.\\

A major surprise in this field were the measurements by Chan and his
collaborators \cite{Moses1,Moses2} on the liquid-vapor transition of helium 
and hydrogen in a variety of porous media. While the liquid-vapor coexistence 
region was considerably shrunk compared to the bulk uniform case,
the exponents were found to be much more akin to those of the uniform
Ising model instead of the RFIM. It has been suggested \cite{Maritan,Maritan2}
that these features are
related to the properties of the RFIM with the fields 
being distributed bimodally
(though not necessarily symmetrically).\\

Another surprise was the novel shape of the experimentally
determined \cite{Moses,Ma,Moses3} 
phase diagram of the $^3$He-$^4$He mixtures
in porous media which allowed for superfluidity at large
concentrations of $^3$He. The classical aspects of this phase
separation are captured by the Blume-Emery-Griffiths (BEG) \cite{BEG}
spin-1 model with an anisotropy.
The presence of a porous medium can be modeled by making this
anisotropy random with a bimodal distribution\cite{Begus}.\\

In this paper, we focus on the role played by
pore walls in liquid-vapor transitions in porous media, as studied
in their corresponding spin $\frac{1}{2}$ Ising spin systems. 
%We demonstrate that the
%effects of confinement are virtually indistinguishable from those 
%predicted for the RFIM with the bimodal distribution of the fields.
There are two aspects to the role of walls in a porous medium.
First, there is a preference for one of the phases over the
other in the vicinity of the walls. This mechanism alone ought to lead
to observable consequences even when the placement of the walls
is substantially periodic, i.e. the different phases are connected
but there is no inherent randomness.
Second, the random placement of the walls in the porous medium
provides quenched disorder and can induce further changes in the
phase behavior. The principal result of our paper is that the former
aspect is more crucial -- indeed, we show that within the
framework of simple models, the phase diagram does not change
on incorporating randomness. This finding is consistent with
the analysis by Galam and Aharony \cite{Galam} indicating that 
the mean field results of a ferromagnet in a random longitudinal
field are the same as a uniaxially anistropic antiferromagnet
in a uniform field. Our results suggest 
that liquid-vapor transitions 
in designed porous media,
with a periodic geometrical pattern \cite{Mallouk,Davis}, 
ought to exhibit behavior quite akin
to that observed in random porous media.
We demonstrate these findings in simple mean field Ising models 
and two distinct values of the local magnetic fields.
We then generalize these studies to spin 1 Ising
models with non-uniform anisotropy and show that such systems behave
like the random anisotropy BEG systems with a bimodal distribution
of the anisotropies \cite{Begus}.
Our results suggest that the $^3$He-$^4$He phase separation
is also primarily governed by the mere presence of walls in the
porous medium and not randomness.\\

It should be noted that there have been several recent mean field
studies of phase transitions in random porous 
media \cite{Kaminsky,De,Kierlik}. One would  expect that
fluctuations in lower dimensions \cite{MacFarland} could play
a key role in qualitatively modifying the mean field picture.\\

\section{Effects of Confinement Random in the Field Ising Model -- 
the Symmetric Case}

We start by considering the simplest case -- that of four Ising spins located
at two kinds of sites, 0 and 1, as shown in 
Figure 1a. Periodic boundary conditions are adopted in the plane
of the figure. Furthermore, it is assumed that above and below this plane, 
there are spins which sit in locations that repeat the pattern shown and
allow for a connected string of nearest neighbor 0 and 1 sites in the
direction perpendicular to the plane of the paper.
Physically, this geometry corresponds to a periodic arrangements of
one-dimensional strings of 0 and 1 arranged on two sublattices.
All spins are coupled by a uniform
exchange constant, $J$. The magnetic field on sites 0 is denoted by $h_0$ and it
points up. On sites 1, on the other hand, the magnetic field is equal to
$h_1$ and it points down. (The case of a simple ferromagnet with
a staggered field is obtained when periodic boundary conditions
are adopted in all directions. One then does not have connected strings of 0 
or 1 sites which lead to an inability to sustain certain phases
at non-zero temperatures.) \\

Our objective here is to determine the phase diagram of this non-random
system within the mean field approximation and compare it to the
corresponding mean field results \cite{Maritan,Maritan2,Aharony} 
of the RFIM in which the probability
distribution of the magnetic fields is bimodal: half of the randomly selected
sites have an up-pointing field $h_0$ and the other half -- a down-pointing
field $h_1$. The RFIM may be thought of as modeling porous media with 
the sites with field $h_0$ corresponding to locations near pore walls and 
the sites  with field $h_1$ to the interior locations. 
The underlying assumption here is that there is
a different environment near the pore wall than
in the interior.\\

The phase diagram is obtained in the three-dimensional
space of $h_0$, $h_1$, and $T$ and is determined by solving the following
equations for the magnetizations $m_0$ and $m_1$:

\begin{equation}
m_0\;=\;tanh[(h_0 + 2Jm_0 + 4Jm_1)/k_BT]  \;\;,
\end{equation}

\begin{equation}
m_1\;=\;tanh[(-h_1 + 4Jm_0 + 2Jm_1)/k_BT]  \;\;,
\end{equation}
on sites with field $h_0$ and $h_1$ respectively. 
The solution is obtained in an iterative
manner that leads to self-consistency. The form of equations (1) and (2)
reflects the fact that each site of a given kind has four neighbors
of the other kind and two neighbors of the same kind -- the latter
resulting from the out-of plane connectivity.
%The phase diagram would look much more simplified if the connectivities 
%between sites of the same kind were eliminated.
Once the solutions for the local magnetizations
are found, one can determine the free energy, $F=U-TS$, by calculating the
internal energy

\begin{equation}
U\;=\;-2 J (m_0 m_0 + m_1 m_1 + 4 m_0 m_1) -2h_0 m_0 + 2 h_1 m_1
\end{equation}

and the entropy 

\begin{equation}
S=2s_0 +2 s_1  \;\;,
\end{equation}

where 

\begin{equation}
s_i=-k_B\frac{1}{2}[(1+m_i)ln(1+m_i)/2 + (1-m_i)ln(1-m_i)/2] \;\;.
\end{equation}
A first order phase transition is identified by the presence of a cusp in the 
free energy.\\

It is easy to show that there are three possible phases at $T$=0 in this system.
We shall denote them by +, $-$, and $+-$ and
their energies by $E_+$, $E_-$, and $E_{+-}$ respectively. In the first phase,
all spins are up and in the second all spins are down. In the third phase,
on the other hand, the spins point in the directions 
of the local magnetic fields.\\

At $T$=0, the + and $-$ phases coexist along the diagonal direction 
in the $h_0 - h_1$ plane (until $h_0=h_1=4J$),
as shown in the top left panel of figure 2.
The $+-$ phase coexists with the + phase
along $h_1=4J$ for $h_0 > 4J$ and with the $-$ phase along $h_0=4J$
for $h_1 > 4J$.
All of the phase boundaries at $T$=0 are first order lines denoted by dashed
lines. The solid lines denote lines of continuous transitions.
Two of these lines occur close to $T=2J/k_B$ and separate 
the $+-$ phase from the
$+$ and $-$ phases respectively. The star in the main phase diagram,
where three critical lines come together, is a tricritical point.
The critical line corresponding to the 
transition between + and $-$  starts
at 6$J/k_B$ when $h_0$ and $h_1$ tend to zero and then decreases steadily as the
fields are increased. In the vicinity of $h_0=h_1$=$\; 4J$, the descent
towards the tricritical point is almost vertical.\\

A particularly simple case is obtained on fixing $h_0$ at the value of $6J$ 
(corresponding to strong pinning at the pore wall)
and varying $h_1$ and $T$
to map out the coexistence curve between the + and $+-$ phases.
In order to cancel the effective field introduced by the nearly
fully aligned $m_0$ spins, one needs to impose a field
$h_1 =4J$ (note that each site '1' has four '0' neighbors)
and effectively one is left with the bulk Ising model.
Figure 3 shows that the coexistence manifests itself in the presence of two
values of $m_1$ but a unique value of $m_0$. 
Of course, a similar scenario takes place
when the boundary between $-$ and $+-$ is crossed.
What is quite remarkable is that the topology of the phase
diagram does not change even when randomness is 
introduced \cite{Maritan,Maritan2,Aharony} in such a way that the symmetry
between the + and $-$ phases is maintained.
Thus within mean field theory, for the symmetric case, the random
placement of the walls plays no role at all. We will show in the
rest of the paper that the same result holds for more complex situations. \\

\newpage

\section{Effects of Confinement in the Random Field Ising Model --  
the Asymmetric Case}

In porous media, the volume of a fluid near the pore walls is usually
much less than the volume of the fluid in the interior. In the Ising spin model,
this translates into an unequal number of sites with fields $h_0$ and $h_1$.
In fact, in a random version of the model, 
we \cite{Maritan} considered
a situation in which a fraction $p$ of the sites has a field $h_0$
-- the symmetric case is obtained when $p=\frac{1}{2}$.

In order to study the effects of walls under such asymmetric conditions,
we consider, for simplicity,
the model shown in figure 1b which is a generalization of figure 1a.
The plane of the figure shows nine sites. The
central site, denoted by 0, has a local field of $h_0$. The remaining eight
sites have a field of $h_1$ and they are denoted by 1 and 2. 
Thus $p=\frac{1}{9}$ but there is no randomness.
The distinction between the two classes of sites, 1 and 2,
is that the former have
site 0 as a neighbor and  the latter do not. Again, it is assumed that above
and below the plane shown there are other planes which repeat the pattern
of the central plane so that each site has a coordination number of six. 
Recall that the boundary conditions along the two directions
within the plane are periodic.\\

The mean field  equations for the three magnetizations read
\begin{equation}
m_0\;=\;tanh[(h_0 + 2Jm_0 + 4Jm_1)/k_BT]  \;\;,
\end{equation}

\begin{equation}
m_1\;=\;tanh[(-h_1 + Jm_0 +3Jm_1 + 2Jm_2)/k_BT] \;\;,
\end{equation}

\begin{equation}
m_2\;=\;tanh[(-h_1 + 2Jm_1 + 4Jm_2)/k_BT] \;\;.
\end{equation}
The internal energy of the system is given by
\begin{equation}
U\;=\;- J (m_0 m_0 + 6 m_1 m_1 + 8 m_2m_2 + 4 m_0 m_1 + 8 m_1m_2) 
-h_0 m_0 + 4 h_1 m_1 + 4 h_1 m_2
\end{equation}
and the entropy by
\begin{equation}
S=s_0 +4 s_1 + 4 s_2 \;\;.
\end{equation}

This system continues to have three phases at $T$=0 as indicated in 
the top left panel of figure 4. The boundaries between the phases, 
however, are shifted to new locations. For instance, the + and $-$ 
phases coexist along the line $h_0=8h_1$, from the origin until $h_0=4J$
The + and $+-$ phases coexist along
$h_1=\frac{1}{2}J$  for $h_0>4J$
whereas the $-$ and $+-$ phases coexist along
$h_0=4J$  for $h_1 > J/2$.
The triple point at which all the three phases coexist is
at $T=0$, $h_0=4J$ and $h_1$=$\frac{1}{2} J$.\\

The emergence of the three phases is the only similarity that
exists between the symmetric and the asymmetric model. The way they 
coexist at non-zero temperatures, for instance, is quite different.
The biggest distinction, shown in the main panel of figure 4, is that now
the two sheets separating the + phase
from the $+-$ 
and the $+$ and $-$ phases along the diagonal direction
combine together to form one surface.
This surface has a tilt that is clearly visible on the
right top panel of figure 4 which shows a section of the phase diagram
at $k_BT=5.5J$. 
The surface terminates at a  critical line
which falls very gently from $k_BT=6J$ at $h_0=0$ 
and $h_1$=0 to about $5.5J$ at $h_0=8.5J$ and $h_1=0.42J$.\\

The coexistence surface of  the $-$ and $+-$ phases continues 
to be substantially planar with a critical line
close to 2$J$ reflecting the one-dimensional
connectivity of the 0 sites. 
This line of critical points intersects the combined
$+,-$ and $+,+-$ coexistence surface at a critical end
point at $k_BT=2.05J$, $h_0=3.9612J$, and $h_1$=$\frac{1}{2}J$.
Quite remarkably,
the topology of this phase diagram is exactly as in the random
case \cite{Maritan}.\\

The coexistence curves for $h_0$=$6J$
(again mimicking a strong pore-wall interaction) are shown in Figure 5. 
Physically, the transition corresponds to crossing from a phase
in which the interior of the pore space is filled by liquid
to one in which the liquid coats the walls and the vapor occupies
the interior. Note the unusual geometry of the coexistence curve.
The magnetizations $m_1$ and $m_2$ have broad 
coexistence curves, similar to
$m_1$ of Figure 3 for the symmetric case.
On the other hand, the coexistence curve for $m_0$ is much narrower
than for $m_1$ and $m_2$ and its non-zero width arises
when the values of $m_1$ and $m_2$ are distinct. When $m_2=m_1$, 
$m_0$ has a unique value in analogy to the symmetric case.
It should be noted that there are just two coexisting solutions --
the larger value of $m_0$ selects positive values of $m_1$ and $m_2$,
whereas the smaller values correspond to negative $m_1$ and $m_2$.\\

\section{Novel Superfluid Phases in $^3$He--$^4$He Mixtures in Aerogel}

We turn now to a discussion of spin systems modeling the phase
separation of $^3$He--$^4$He mixtures.
Figure 6(a) shows a sketch of the experimental phase diagram
(in the temperature ($T$) - concentration of $^3$He ($x$) plane) of bulk
$^3$He-$^4$He mixtures in the vicinity of the 
superfluid transition of $^4$He.  In
the temperature range of interest, the superfluid transition involving the
pairing of $^3$He atoms is not a factor and  indeed 
the $^3$He atoms can be thought
of as inert, annealed (i.e. they are not stuck in space but can move
around) entities.  At low $^3$He concentrations, on cooling the system, a
superfluid transition denoted by the solid line (AB) is observed.  However
at higher $^3$He concentrations, the system 
opts to phase separate into a $^4$He
rich region which becomes superfluid.
The coexistence curve of the $^3$He-$^4$He
phase separation is shown as a dashed curve (CBD).  Two interesting
features of the phase diagram are the tricritical point B, where the
superfluid transition line collides with the coexistence curve at its
critical point and the miscibility gap at C -- small amounts of $^3$He 
added to $^4$He do not lead to phase separation: 
a feature exploited in dilution
refrigerators.\\

Perhaps, the simplest classical model that captures the topology
of this phase diagram is the Blume-Emery-Griffiths model \cite{BEG} (BEG) 
which is a lattice model
populated with spins, $S_i$, that can take on one of three values
0, -1, or 1.  The inert $^3$He is represented by 0 spins 
and $^4$He is denoted by
+1 or 1 spin values.  An exchange coupling between nearest neighbor
non-zero spins, favoring alignment causes the analog of the superfluidity
transition, with the broken symmetry phase having a non-zero magnetization
(i.e. a mismatch in the number of +1 and 1 spins).  The Hamiltonian reads
\begin{equation}
H\;=\; -J \sum _{<ij>} S_iS_j \;+\; \sum _i \Delta S_i^2 \;\;,
\end{equation}
where $\Delta $ is an anisotropy field which controls the relative
concentrations of the two isotopes.
The presence of $^4$He corresponds to $S_i=\pm1$,
superfluidity of $^4He$ to the existence of non-zero magnetization
and the $^3$He atoms are represented by $S_i=0$. 
The random anisotropy field here
does not break the $\pm$ symmetry.\\

The resulting phase
diagram (shown in Figure 6b) has all the correct qualitative features,
except for the absence of the miscibility gap at C which is believed to
arise from a purely quantum mechanical effect.  Even thought the BEG model
is purely classical and does not have the correct symmetry of the spins
(the superfluid transition has the same characteristics of the transition
in a $xy$ model in which spins lie in a plane rather than having up-down
symmetry), it nevertheless reproduces almost all the qualitative features
of the experiment correctly.\\

Recent experiments of Chan and 
coworkers \cite{Moses,Ma,Moses3}
on the phase separation of $^3$He-$^4$He
mixtures in aerogel in the vicinity of the superfluid transition
have yielded a phase diagram shown Figure 6(c).  The key features of the
phase diagram are i) the absence of the tricritical point (the superfluid
transition line no longer intersects the coexistence curve); ii) an
{\em enhancement} of the superfluid transition temperature compared to the bulk
at large $^3$He concentration iii) at low temperatures (below the critical
point associated with the phase separation) and for a range of values of $x$
within the coexistence curve, 
$^4$He rich and $^3$He rich regions coexist, {\em both}
of which are superfluid and iv) the experimental data, while restricted to
temperatures above 0.35 K, are suggestive that the aerogel causes a
miscibility gap to open up at large value of $x$.  This is of fundamental
importance, if true, since the superfluid phase observed is one in which a
small quantity of $^4$He in $^3$He does not phase separate (as observed in the
bulk), but is yet superfluid and probably represents the long sought after
dilute Bose gas superfluid phase.  Even more exciting, such a miscibility
gap would lead to the extremely novel situation of two distinct coexisting
superfluid phases at low temperatures, 
the dilute Bose gas phase of $^4$He and the superfluid
phase of $^3$He.  Two factors in support of the novel dilute superfluid
phase are: a) Adding a small amount of $^4$He to the aerogel (in
the absence of $^3$He) leads to a superfluid phase 
whose density is {\em enhanced}
by the addition of $^3$He. b) Because a coexistence curve for the phase
separation is found in the phase diagram, 
it is plausible that there is {\em no} phase
separation in the region between B and D (6(c)), since it is unlikely
to expect phase separation of an already phase separated phase.\\

\vspace*{-1cm}
\section{The Random Anisotropy BEG Model}
%$P_0\;=\;1-q$, $P_1\;=\;(q+m)/2$, $P_{-1}\;=\;(q-m)/2$

In the BEG model the effect of the aerogel is assumed to be
present on a fraction p of the sites, these sites are randomly chosen and
fixed (unlike the mobile $^3$He atoms, the aerogel is a manifestation of
quenched randomness). The distribution of the single site
anisotropies is bimodal and given by \cite{Begus}
\begin{equation}
P(\Delta _i )=\;=\; p\delta(\Delta _i -\Delta _0) \;+\; 
(1-p)\delta (\Delta _i - \Delta _1) \;\;.
\end{equation}
The sites with anisotropy $\Delta _0$ correspond
to vicinity of pore walls and for the situation in which
$^4$He prefers to be near the wall,
$\Delta _0 < 0$. The $\Delta _1$ anisotropy characterizes the pore
interior and its value controls the total number of $^3$He atoms.
In a mean field approach, one obtains the phase
diagram shown in Figure 6(d).  Note that this is in accord with the
experimental observations (i) and (iii), but does not reproduce (ii) and
(iv).  The tricritical point (where three phases go critical
simultaneously) requires a special symmetry, which is absent when one
incorporates the random anisotropy to mimic the
aerogel.  The line of superfluid transitions is, however, virtually
insensitive to the presence of the random anisotropy.  The coexistence curve, 
however, is shifted to lower temperatures and higher effective $^4$He 
concentration due to the space taken up by the aerogel, thus leading to the
topology shown in Figure (6(d)).\\

In order to investigate whether this lack of complete agreement
arises due to quantum mechanical effects and their neglect by the BEG model
or due to the inherent simplicity of the mean field approach, we have 
carefully studied the BEG model within an improved mean field
theory which is a generalization of the approach presented
in Ref. \cite{Improved}. The improved method captures features
such as a percolation threshold and yield better estimates
of the transition temperature than the mean field theory.\\
%two different approaches that are more
%sophisticated than the mean field theory.  Our analysis has been carried
%out using an improved mean field theory and a cluster variational method,
%both of which capture features such as a percolation threshold and yield
%better estimates of the transition temperature than mean field theory.
%The improved mean field theory is a generalization of the approach
%presented in Ref. \cite{Improved}.
%The approximation scheme of the cluster variation 
%method \cite{Kikuchi,Kikuchi1} is determined by the largest clusters
%of sites that are treated exactly. The scheme generates
%an expression for the free energy of the system as a function
%of the probability of occurrence of all possible configuration
%of the basic cluster. This free energy is then minimized subject
%to consistency conditions on the distribution variables.\\

  The phase diagrams 6(e) and (f) are obtained depending on whether
the fraction $p$ of the sites at which the  aerogel is present is less
than or greater than the percolation threshold $p_c$.  Unlike the porous
medium aerogel, which has a strongly correlated, connected interface, our
model, in its simplest form, consists of randomly chosen interface sites
allowing for a percolation threshold.  In the experiment, in spite of the
large porosity, one is always in the fully connected regime.  Note the
presence of a miscibility gap at high $^3$He concentrations.  Unlike mean
field theory (6(d)), the point D is at a concentration less than
$(1-p)$.  Indeed, our calculations suggest a second coexistence curve
between D and the point ($x=1-p$; $T$=0) analogous to
the one between C and D except that the
critical temperature is shifted down to zero.  Thus for concentrations of
$^3$He corresponding to points between D and ($x=1-p$; $T$=0), 
and temperatures less than the
superfluid transition line AB, the model predicts the analog of the dilute
Bose gas superfluid phase.\\  

The superfluid transition temperature smoothly extrapolates
to the value of the transition temperature of a coated phase of helium
atoms residing on the aerogel surface.  This is dramatically seen in
Figure 6(f) where the transition plunges to zero at $x = 1-p$ when $p$ is
less than the percolation threshold $p_c$ and is unable to sustain
a phase transition at non-zero temperatures.  Within the
context of the BEG model, the novel superfluid phase is found to be one in
which the magnetization (superfluidity) arises from the aerogel sites and
from the sites in their vicinity.  Indeed, in any classical model with
short range interactions, the spins yielding a non-zero magnetization must
lie on a connected cluster and are thus in an essentially phase separated
phase.  This phase separation, however, does {\em not} preclude a further
bulk-like phase separation on increasing the concentration of $^4$He
atoms.  In our BEG model, the minimum number of $^4$He atoms is equal to
the number of aerogel surface sites.  A further addition of $^4$He 
atoms (in the absence
of $^3$He) causes an increase in the magnetization, corresponding to
the attachment of some of these atoms to the already existing spanning
cluster at the aerogel surface.  Subsequent addition of $^3$He atoms results
in more of the $^4$He atoms going in the cluster, thereby enhancing the
magnetization, as in the experiment.  We have also studied the effects of
correlation in the selection of aerogel surface sites: the probability of a
nearest neighbor site of an aerogel surface site to be another aerogel
surface site is enhanced compared to a completely random selection.  We
find this correlation enhances the superfluid transition temperature
compared to the bulk in accord with experiment.\\

In summary, a simplified model for $^3$He-$^4$He mixtures in aerogel
reproduces many (but the miscibility gap at low $^3$He concentrations) 
of the features
observed in experiments and suggests the opening of a
miscibility gap at low $^4$He concentration.  The analog of the dilute Bose
gas phase within the classical model is one in which the superfluidity
arises from $^4$He adsorbed on the aerogel.  Note that this does not however
preclude further phase separation.  Cooling the system to ultralow
temperatures in the miscibility gap at high $^3$He concentrations should lead
to two coexisting superfluid phases: $^3$He and the $^4$He near the aerogel.
It would be very exciting if quantum mechanical effects (not
considered here) delocalize the $^4$He atoms leading to a dilute Bose gas
phase at higher temperatures and interpenetrating $^4$He and $^3$He 
superfluid phases at low temperatures.\\

It should be noted that much of the physics corresponding to the
scenarios of Figure 6 has been captured by the renormalization
group analysis of Berker and his 
collaborators \cite{Berker,Berker1,Berker2,Berker3}.
They considered random and non-random
"jungle-gym" models of the aerogel and explained
the phase diagrams
by the connectivity and tenuousness of the aerogel.\\

\section{Effects of Confinement in the BEG model}

In order to study the separate roles of the presence of walls
and randomness on $^3$He--$^4$He separation,
we again consider the geometry shown in Figure 1b and set up
mean field equations that correspond to the spin-1 problem.
These equations for the magnetizations, i.e. the expectation values of
$S_i$, and for the three parameters $q_0$, $q_1$, and $q_2$ which
are the expectation values of $S_i^2$, are:

\begin{equation}
m_0\;=\;q_0 \; tanh[(2Jm_0 + 4Jm_1)/k_BT] \;\;,
\end{equation}
\begin{equation}
m_1\;=\;q_1 \; tanh[(Jm_0 +3Jm_1 + 2Jm_2)/k_BT] \;\;, 
\end{equation}
\begin{equation}
m_2\;=\;q_2 \; tanh[(2Jm_1 + 4Jm_2)/k_BT] \;\;,
\end{equation}
and
\begin{equation}
q_i\;=\;\frac{4\; \exp{(-2\Delta _i/k_BT)} - \sqrt{m_i^2  +
 4 \; \exp{(-2\Delta_i /k_BT)} \;
(1-m_i^2)}}   {4\; \exp{(-2 \Delta _i /k_BT)} -1} \;\;.
\end{equation}
The internal energy is given by
\begin{equation}
U\;=\;- J (m_0 m_0 + 6 m_1 m_1 + 8 m_2m_2 + 4 m_0 m_1 + 8 m_1m_2) 
+\Delta_ 0 q_0 + 4 \Delta_ 1 q_1 + 4 \Delta_ 1 q_2
\end{equation}
and the entropy by
\begin{equation}
S=\tilde{s}_0 +4 \tilde{s}_1 + 4 \tilde{s}_2 \;\;,
\end{equation}
where
\begin{equation}
\tilde{s}_i=-k_B[(1-q_i) ln(1-q_i) +
\frac{1}{2}(q_i + m_i)ln(q_i + m_i)/2 + (q_i - m_i)ln(q_i - m_i)/2] \;\;.
\end{equation}

\vspace*{0.5cm}

In direct analogy to the random case \cite{Begus}, 
there are four posssible phases at $T$=0:
\begin{description}
\item {\bf phase 1} in which all three $m_i$s are non-zero;
\item {\bf phase 2} in which $m_0 > 0$, $m_1=m_2=0$;
\item {\bf phase 3} in which $m_0=0$, $m_1$, $m_2 > 0$; and
\item {\bf phase 4} in which $m_0=m_1=m_2=0$. 
\end{description}
In each phase,
$q_i$=$m_i$ at $T$=0. The non-zero magnetization
persists to higher temperatures
and its disappearance corresponds to the $\lambda$ line
for superfluid $^4$He. (Note that our analysis lumps in any inert
or dead layer of $^4$He as belonging to the pore wall.)
The analog of the $^3$He concentration is given by
\begin{equation}
x\;=\; 1 \;-\; (q_0 + 4q_1 + 4 q_2)/9 \;\;.
\end{equation}

The overall topology of the phase diagram is shown on figure 7 and
several isothermal slices through it are shown in figure 8. At low
temperatures all four phases exist and their number goes down
on increasing the temperature.
The $T$=0 boundaries are given by
\begin{description}

\item 1 -- 2 coexistence at $\Delta _1=\frac{13}{4}J$, $\; \Delta _0 < J$;

\item 1 -- 3 coexistence at $\Delta _0=5J$, $\; \Delta _1 < 
\frac{11}{4} J$;

\item 1 -- 4 coexistence 
$\Delta _1=\frac{27}{8}J - \frac{1}{8} \Delta _0$, $\; J < \Delta _0 < 5J$ ;

\item 2 -- 4 coexistence at $\Delta _0 = J$, $\; \Delta _1 > \frac{13}{4}J$;

\item 3 -- 4 coexistence at $\Delta _1 =\frac{11}{4} J$, 
$\; \Delta _0 > 5J$.

\end{description}
These first-order boundaries become vertical surfaces 
on considering the $T$-axis. The top edges of these surfaces
are critical lines. 
This has as its roof
a critical surface at which the magnetization
disappears. Of course phase 4, which is paramagnetic, is not covered
by a roof.
This roof corresponds to the 
superfluid transition of $^4$He  or the $\lambda$ - line.
The shape pf the roof is illustrated in figure 9 for several
values of $\Delta _0$. 
In the figure, for $\Delta _0 /J \le 0.8$, the roof continues
indefinitely for large $\Delta _1$, becaouse $m_0$ remians zero at
sufficiently low temperatures. However, when $\Delta _0/J$, 
the roof smoothly  terminates at the top of the wall separating
planes 1 and 4. This is necessitated by the fact that phase 4 has no
roof.\\

Figure 10 illustrates the nature of the phase diagram
for selected values of $\Delta _0$. The insets show the transition
lines as a function of $\Delta _1$ and the main figures -- as
a function of $x$ -- the analog of the $^3$He concentration.
The top two panels of Figure 10 refer
to the uniform anisotropy case - when
$\Delta _0 $ is equal to $\Delta _1$ and confirm that
this simple nine-spin model captures the topology of the phase
diagram of the uniform BEG model \cite{BEG}.\\

The physically interesting regime is that of negative $\Delta _0$
which favors $^4$He near the pore walls. It is seen
that, as $\Delta _1$ is varied, the $\lambda$ line becomes disconnected
from the phase separation coexistence lines (the middle panels).
The details depend on how one moves on the $\Delta _0$ - $\Delta _1$
plane. For instance, if one crosses the 1--2 and 2--4 
boundaries at an angle, one gets the situation depicted in 
the bottom panels of Figure 10.
All of these phase diagrams are 
in accord with the random anisotropy version of 
the model except that the $\lambda$ line of the bottom panels of
figure 10 does not reemerge from the  $^3$He-$^4$He coexistence region
because, this region ends at $x$=1 in this simple model, and not at an 
$x$ which is less than 1.
Figure 11 shows the coexistence curves for the $q$
order parameters for $\Delta _0$=-0.5$J$. They are remarkably similar
to the magnetization coexistence curves of Figure 6. In particular, 
the non-zero width for the coexistence region in $q_0$ reflects
inequality of $q_1$ and $q_2$.\\

The basic message of our analysis is that the topology of the phase
diagram changes qualitatively (in accord with experiment) when one
moves in the $(\Delta _0 - \Delta _1)$ plane such that instead of
going directly from phase 1 to the paramagnetic phase 4, one
moves through the intermediate phase 2.
Somewhat surprisingly and at odds with expectations, one finds that
the mean field topology of the phase diagram  is more sensitive to
the presence of walls in a porous medium than to the role played
by their random placement. A similar conclusion has been reached by
Pricaupenko and Treiner \cite{Pricaupenko} within a nonlocal
density functional analysis of $^3$He-$^4$He mixtures in a channel
geometry. In particular  their analysis shows possibility of the detachment
of the superfluid line from the coexistence region.
It would be interesting to consider whether fluctuations make a 
qualitative difference in the conclusions reached in our simple mean field
analysis.\\

We are indebted to Moses Chan for stimulating discussions.
This work was supported by
the Center of Collective Phenomena in
Restricted Geometries, the Penn State MRSEC under NSF grant DMR-0080019,
INFM, %KBN (Grant No. 2P03B-146-18), 
MURST, and NASA.

\newpage
FIGURE CAPTIONS
\begin{description}

\item Figure 1. a) The basic unit of the model used to study 
the symmetric case 
in which there are as many sites with the local field $h_0$, denoted by 0,
as with the field $h_1$, denoted by 1.
The 0s and 1s are placed on two sublattices in the plane as shown and
is periodically repeated in both directions in the plane of the paper.
The model has lines of 0 and 1 sites respectively perpendicular
to the plane of the paper. 
In other words, the pattern shown is repeated in other parallel planes.
b) The geometry of the model used to study the 
asymmetric case.
The site denoted by 0 has a field $h_0$ and the remaining
sites a field $h_1$.
As before, there are periodic boundary conditioons in the plane
and repeat boundary conditions in the direction perpendicular to the
page.

\item Figure 2. The main panel shows the phase diagram 
corresponding to the
symmetric case in a three dimensional representation $h_0 - h_1 - T$.
The dashed lines correspond to first order transitions whereas 
the thick solid lines correspond to the continuous transitions.
The star indicates a tricritical point.
The top panels show constant temperature slices of the phase diagram
for the temperatures indicated. The asterisk on the right hand panel 
indicates a critical point and the square on the top left
panel is a triple point. 

\item Figure 3. The temperature dependence
of the local magnetizations, $m_0$ and $m_1$
on crossing the boundary between the + and $+-$ phases at $h_0=6J$. 
For temperatures below $T=2J$, there is a coexistence of two values of
$m_1$ but $m_0$ stays essentially fully magnetized up to $T\;=\; 2J$.

\item Figure 4. The main panel shows the phase 
diagram corresponding to the
asymmetric case in a three dimensional representation $h_0 - h_1 - T$.
The dashed lines correspond to first order transitions whereas
the thick solid lines correspond  to continuous transitions.
The pentagon indicates a critical end point.
The top panels show constant temperature slices of the phase diagram
for the temperatures indicated. 
The asterisk on the right hand panel 
indicates a critical point, 
whereas the square on the top left
hand panel is a triple point.

\item Figure 5. The temperature dependence
of the local magnetizations, $m_0$, $m_1$, and $m_2$
on crossing the boundary between the + and $+-$ phases at $h_0=6J$. 
The coexistence curve for $m_0$ is much narrower than for $m_1$ and
$m_2$. There are two coexisting solutions and the larger value of
$m_0$ selects positive values of $m_1$ and $m_2$, whereas the smaller values
correspond to negative $m_1$ and $m_2$.

\item Figure 6.
Schematic representations of the experimentally (panels a and c)
and theoretically determined phase diagrams for $^3$He-$^4$He
separation as described in more detail in the text. 
The phase diagram is shown in the temperature ($T$) -- concentration
of $^3$He ($x$) plane.
Panels
a) and b) correspond to experimental and theoretical results
for the bulk case. All other panels refer to random
situations. Panel c) corresponds to the experiments in aerogels.
Panel d) corresponds to the mean field analysis of the random
anisotropy BEG model with a bimodal distribution of the anisotropies.
Panels e) and f) correspond to theoretical results in which
$p$, the fraction of randomly chosen sites corresponding to the pore walls,
is smaller or larger than the percolation threshold respectively.

\item Figure 7. The phase diagram for the BEG model with a bimodal
non-random 
distribution of the anisotropies in a three-dimensional representation
$\Delta _0 - \Delta _1 - T$.
The broken lines correspond to the first order transitions whereas 
the solid lines correspond to continuous transitions.

\item Figure 8.
Constant temperature slices of the phase diagram shown in Figure 7
at the temperatures indicated. 

\item Figure 9. Plot of the critical lines (the temperature at which the
magnetization goes to zero) as a function of $\Delta _1$ for the 
indicated values of
$\Delta _0$.

\item Figure 10. {\bf Top left panel:}
The phase diagram for the uniform (bulk) anisotropy BEG
model in the $T - x$ plane, where $x$ is the spin analog of the
concentration of He$^3$. {\bf Top right panel:}
The same phase diagram but in the $T - \Delta _1$ plane.
$\lambda$  indicates the critical 
line for the
magnetization and I the first order transition in the 
$q$-order parameter.
{\bf Middle left panel:}
The phase diagram for the non-uniform (periodic) anisotropy BEG
model in the  $T - x$ plane. The data are for $\Delta _0\;=\;-0.5J$.
{\bf Middle right panel:}
The corresponding phase diagram in the $T - \Delta _1$ plane.
{\bf The bottom panels:} The phase diagrams for
for $\Delta _0 = \Delta _1 /5$.

\item Figure 11. The coexistence curves for the order parameter $q$
on the three sites when $\Delta _0 = -0.5 J$.
There are two coexisting solutions and the larger value of
$q_0$ selects larger values of $q_1$ and $q_2$.

\end{description}

\newpage
%FIGURE 1
\begin{figure}
%\vspace*{-5.5cm}
\epsfxsize=4.5in
\hspace*{-0.5cm}
\centerline{\epsffile{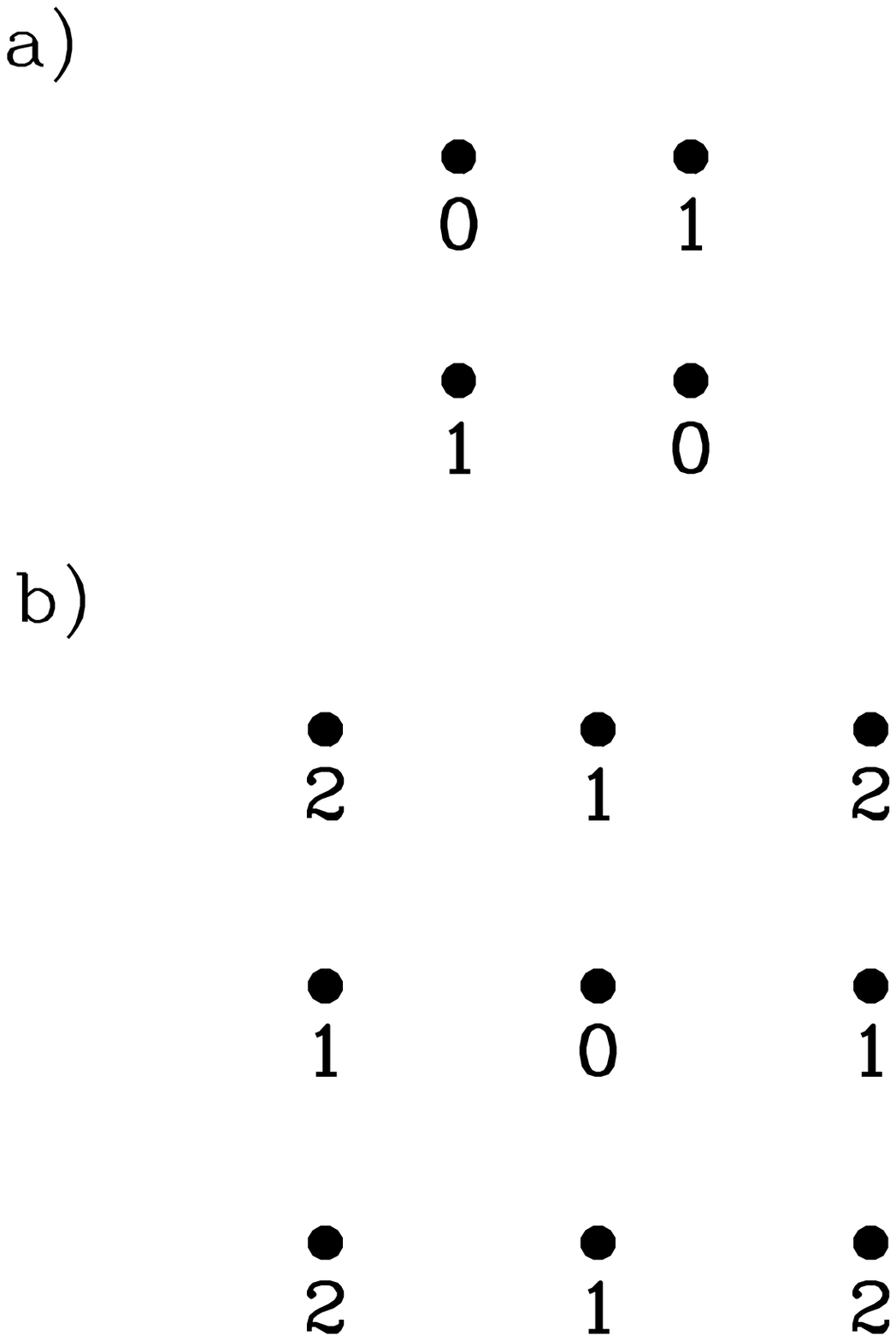}}
%\vspace*{1.8cm}
\caption{ }
\end{figure}

%FIGURE 2
\begin{figure}
\vspace*{1.5cm}
\epsfxsize=4.5in
\hspace*{-0.5cm}
\centerline{\epsffile{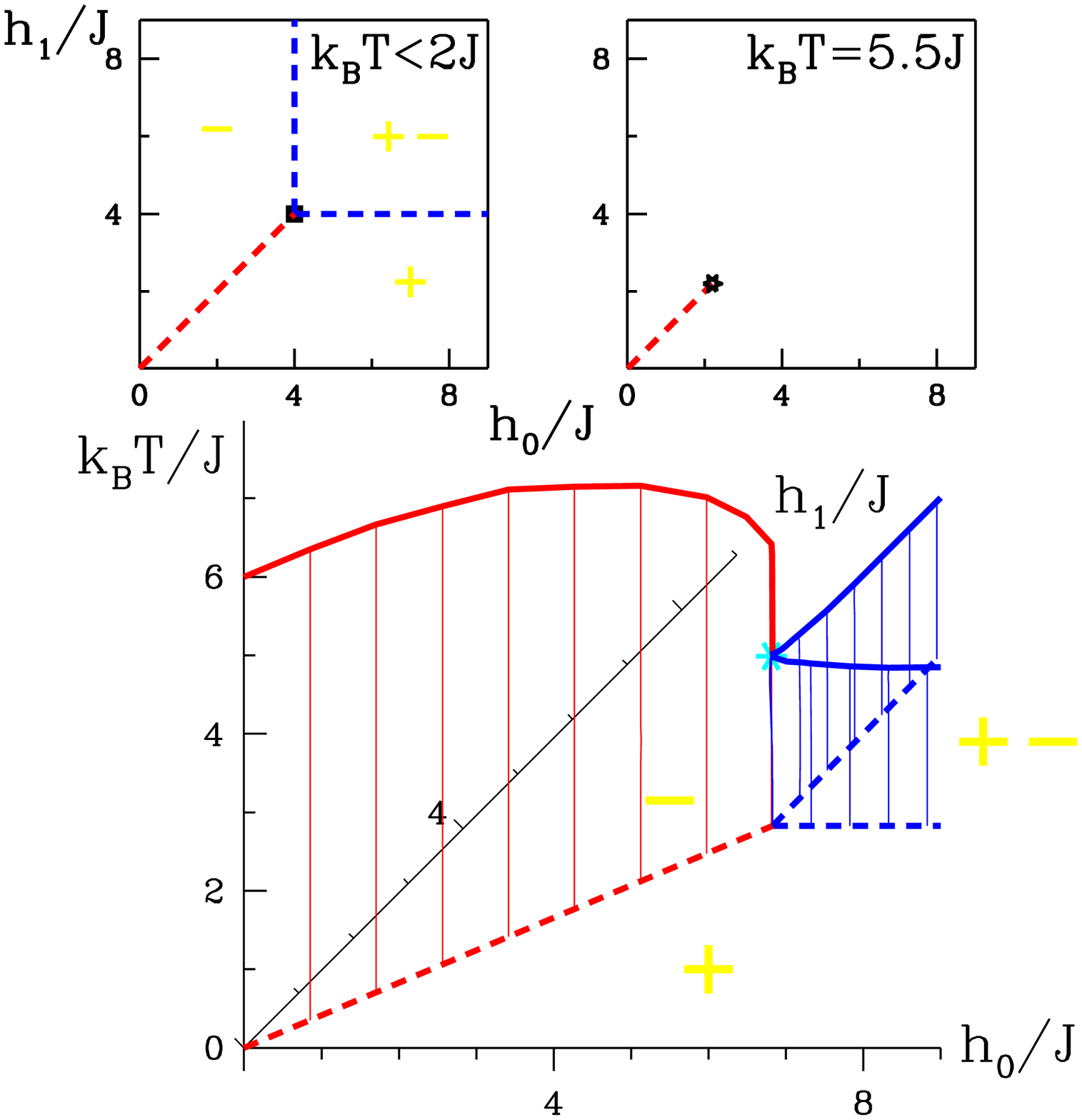}}
%\vspace*{1.8cm}
\vspace*{2.3cm}
\caption{ }
\end{figure}

%FIGURE 3
\begin{figure}
\vspace*{0.5cm}
\epsfxsize=5.0in
\hspace*{-0.5cm}
\centerline{\epsffile{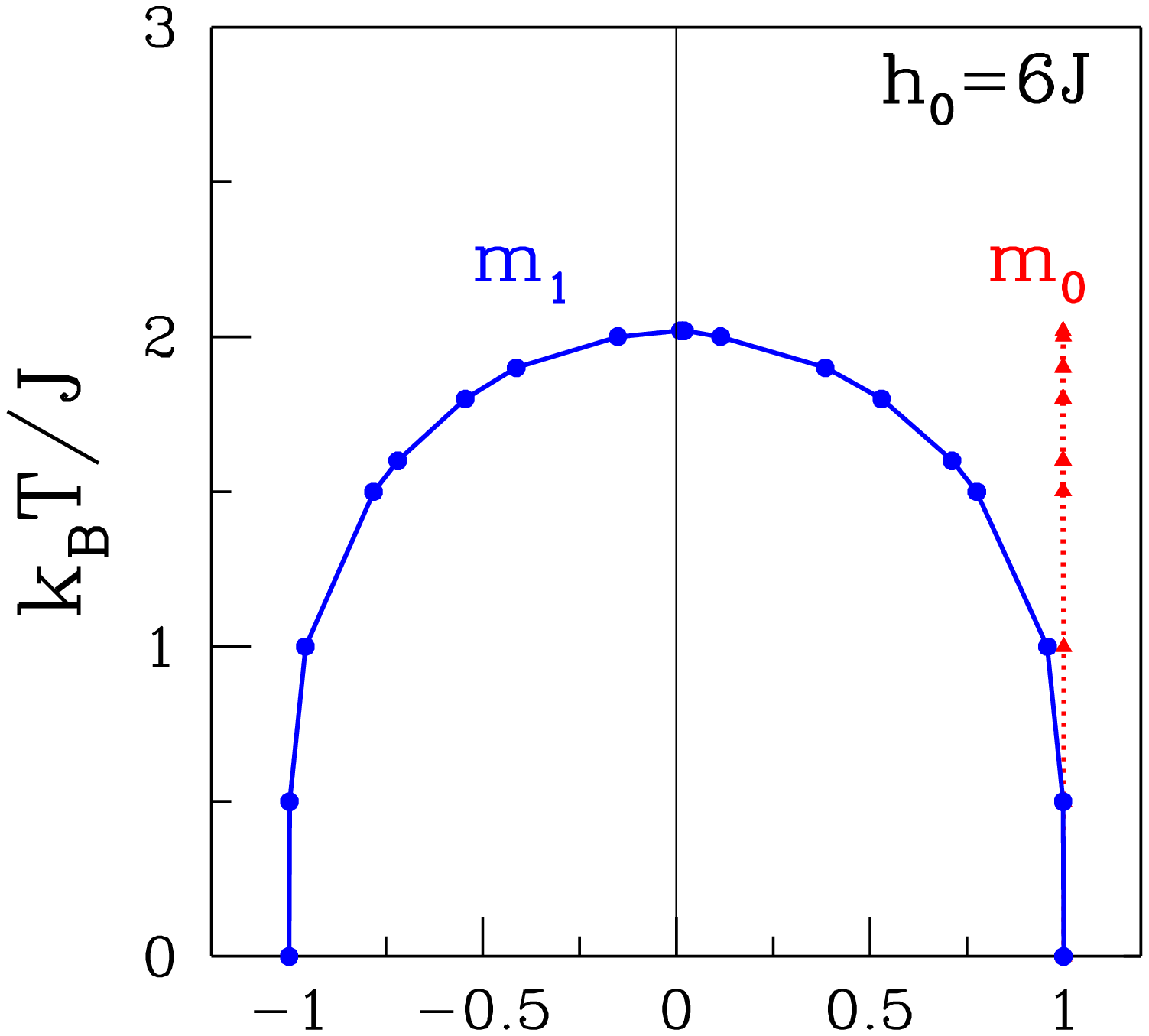}}
%\vspace*{1.8cm}
\vspace*{2.3cm}
\caption{ }
\end{figure}

%FIGURE 4
\begin{figure}
\vspace*{1.5cm}
\epsfxsize=4.5in
\hspace*{-0.5cm}
\centerline{\epsffile{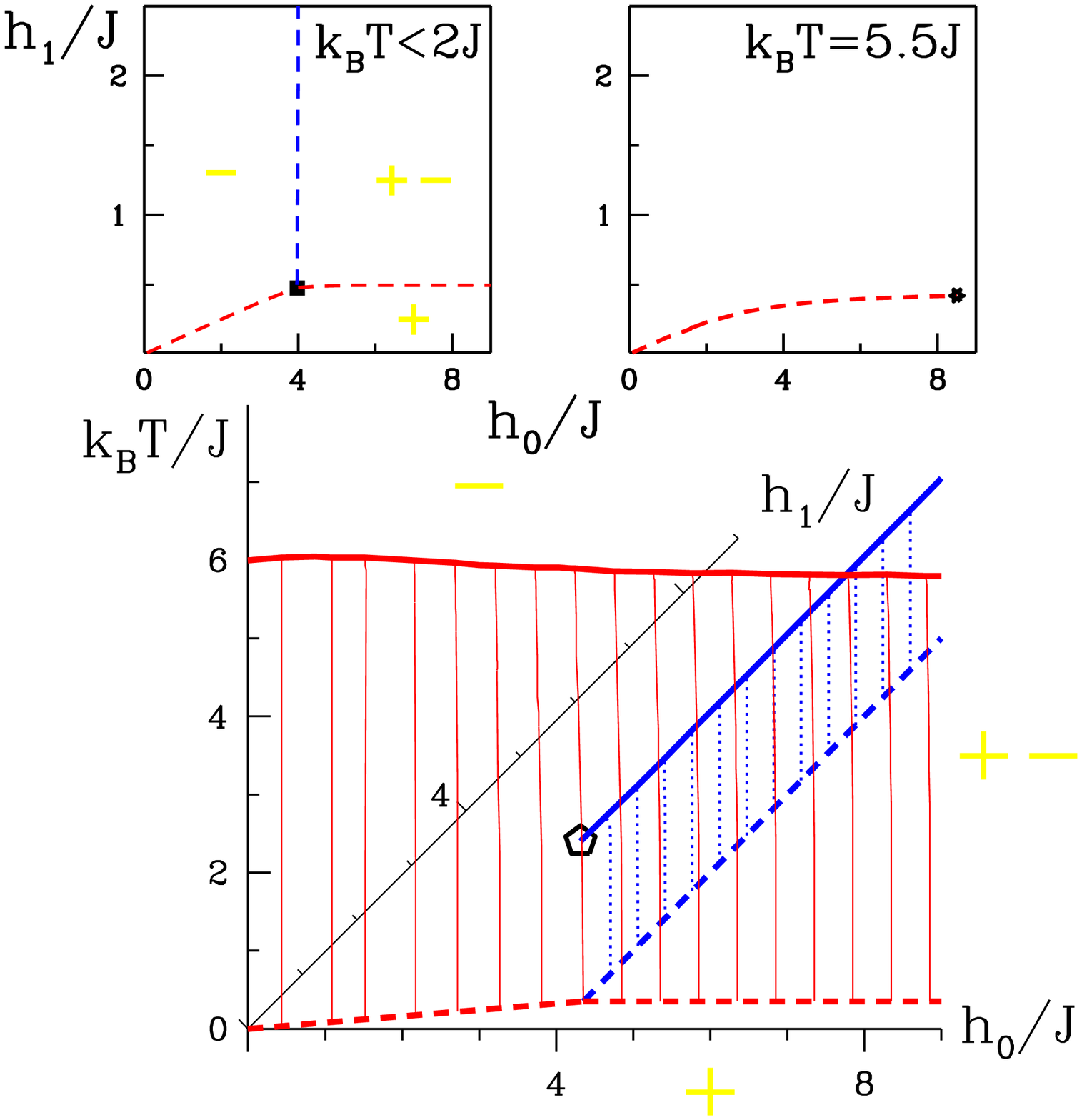}}
%\vspace*{1.8cm}
\vspace*{2.3cm}
\caption{ }
\end{figure}

%FIGURE 5
\begin{figure}
\vspace*{0.5cm}
\epsfxsize=5.0in
\hspace*{-0.5cm}
\centerline{\epsffile{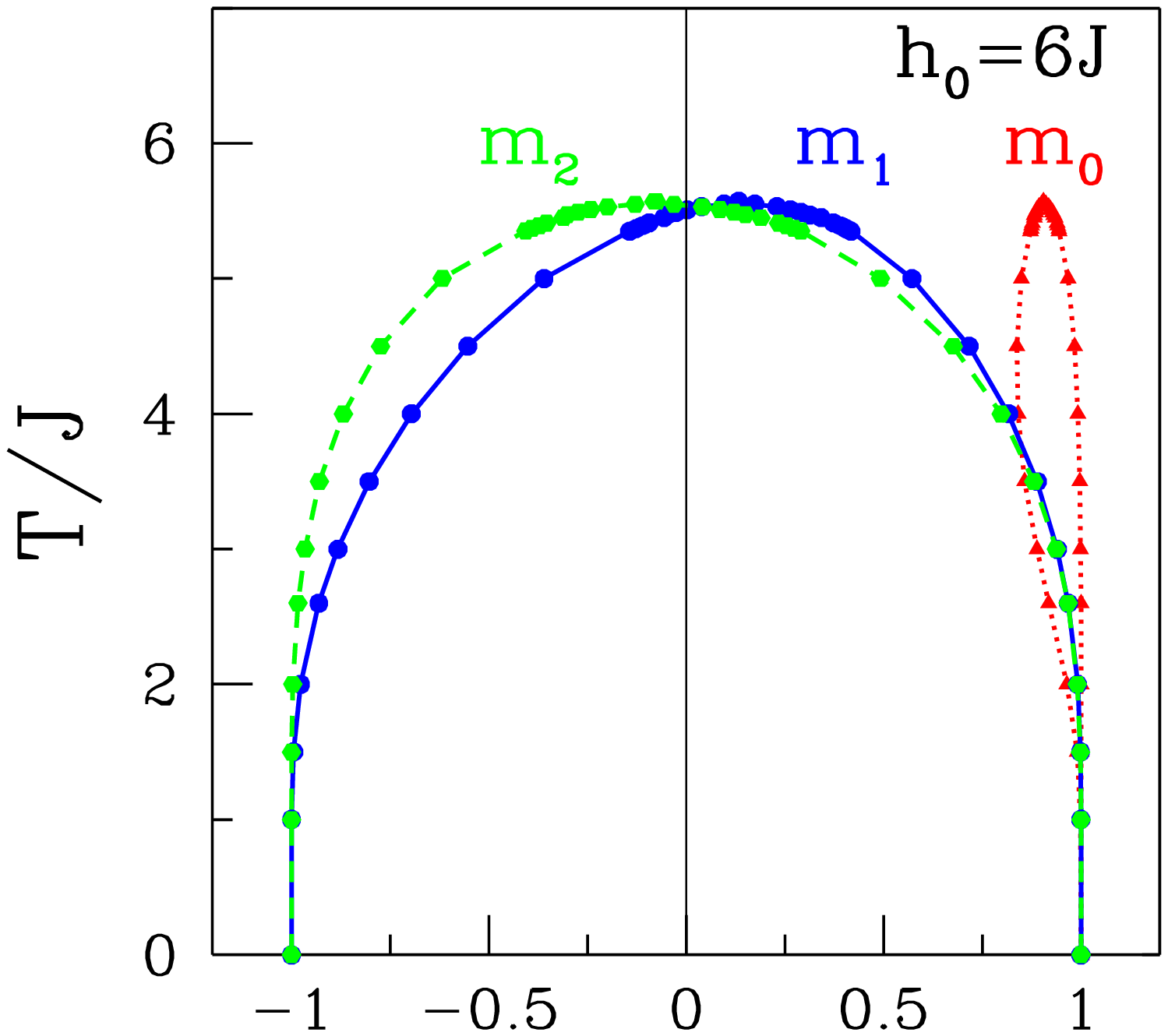}}
%\vspace*{1.8cm}
\vspace*{2.3cm}
\caption{ }
\end{figure}

%FIGURE 6
\begin{figure}
\vspace*{1.5cm}
\epsfxsize=4.5in
\hspace*{-0.5cm}
\centerline{\epsffile{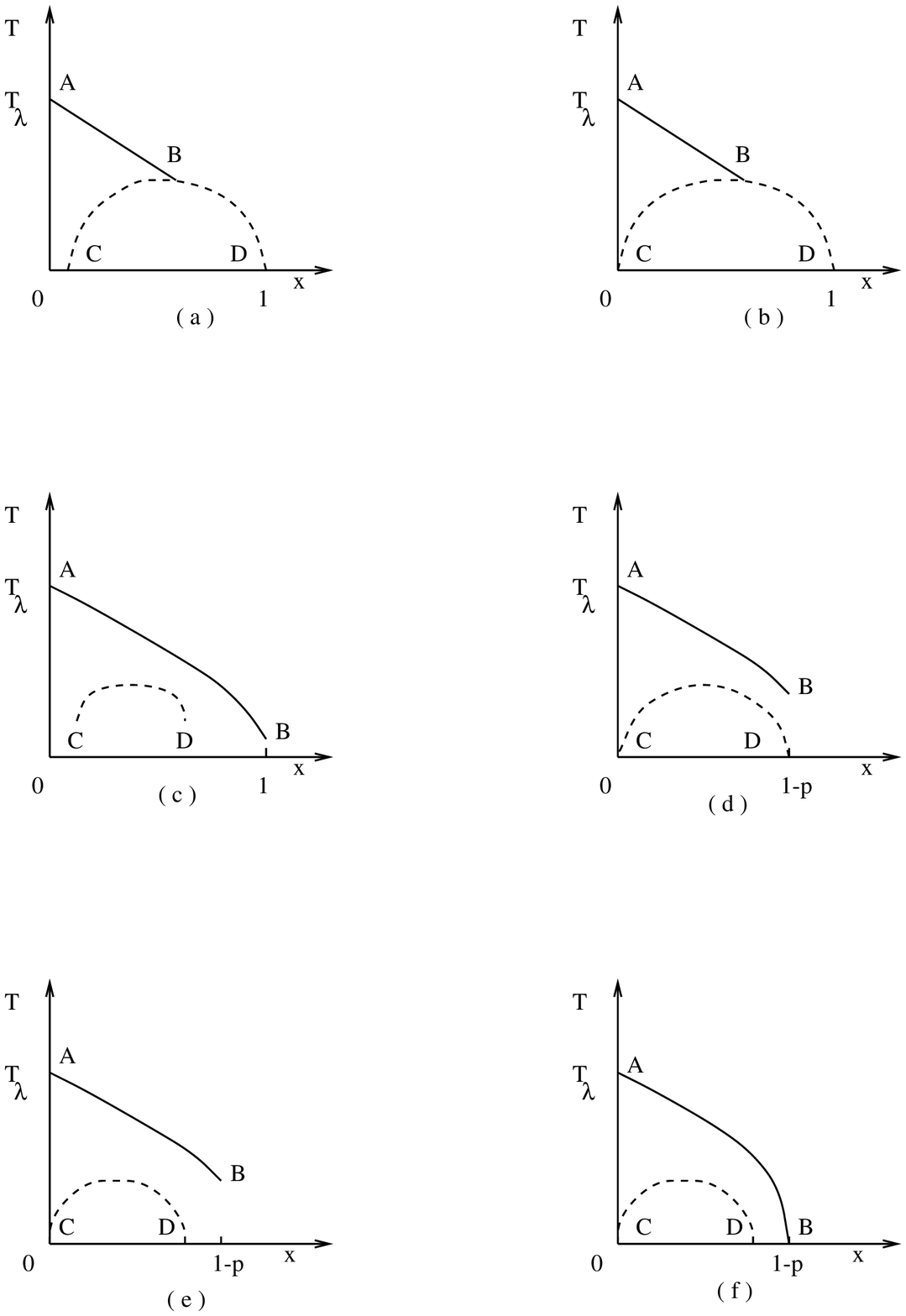}}
%\vspace*{1.8cm}
\vspace*{1.3cm}
\caption{ }
\end{figure}

%FIGURE 7
\begin{figure}
\vspace*{1.5cm}
\epsfxsize=4.5in
\hspace*{-0.5cm}
\centerline{\epsffile{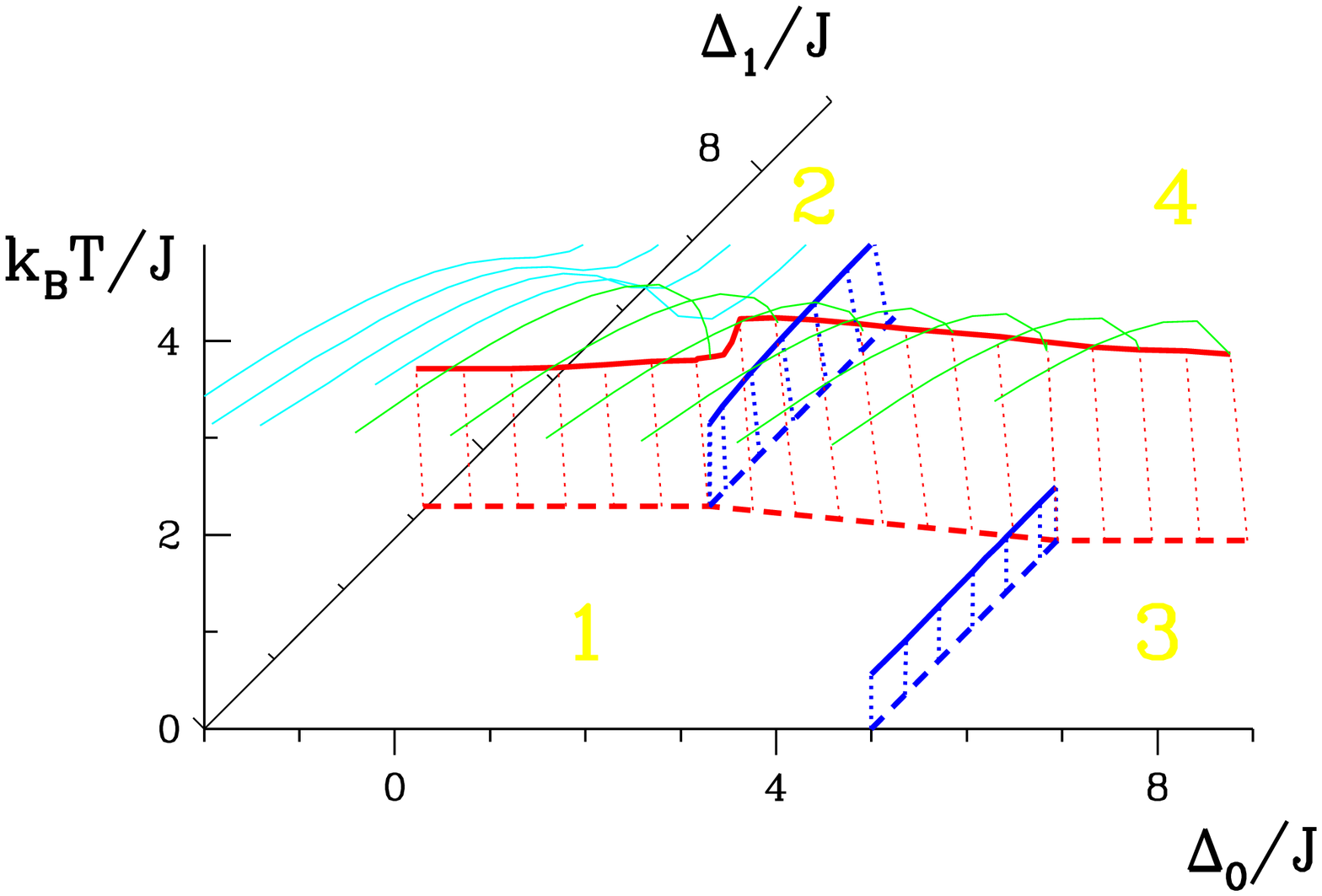}}
%\vspace*{1.8cm}
\vspace*{2.3cm}
\caption{ }
\end{figure}

%FIGURE 8
\begin{figure}
\vspace*{1.5cm}
\epsfxsize=5.0in
\hspace*{-0.5cm}
\centerline{\epsffile{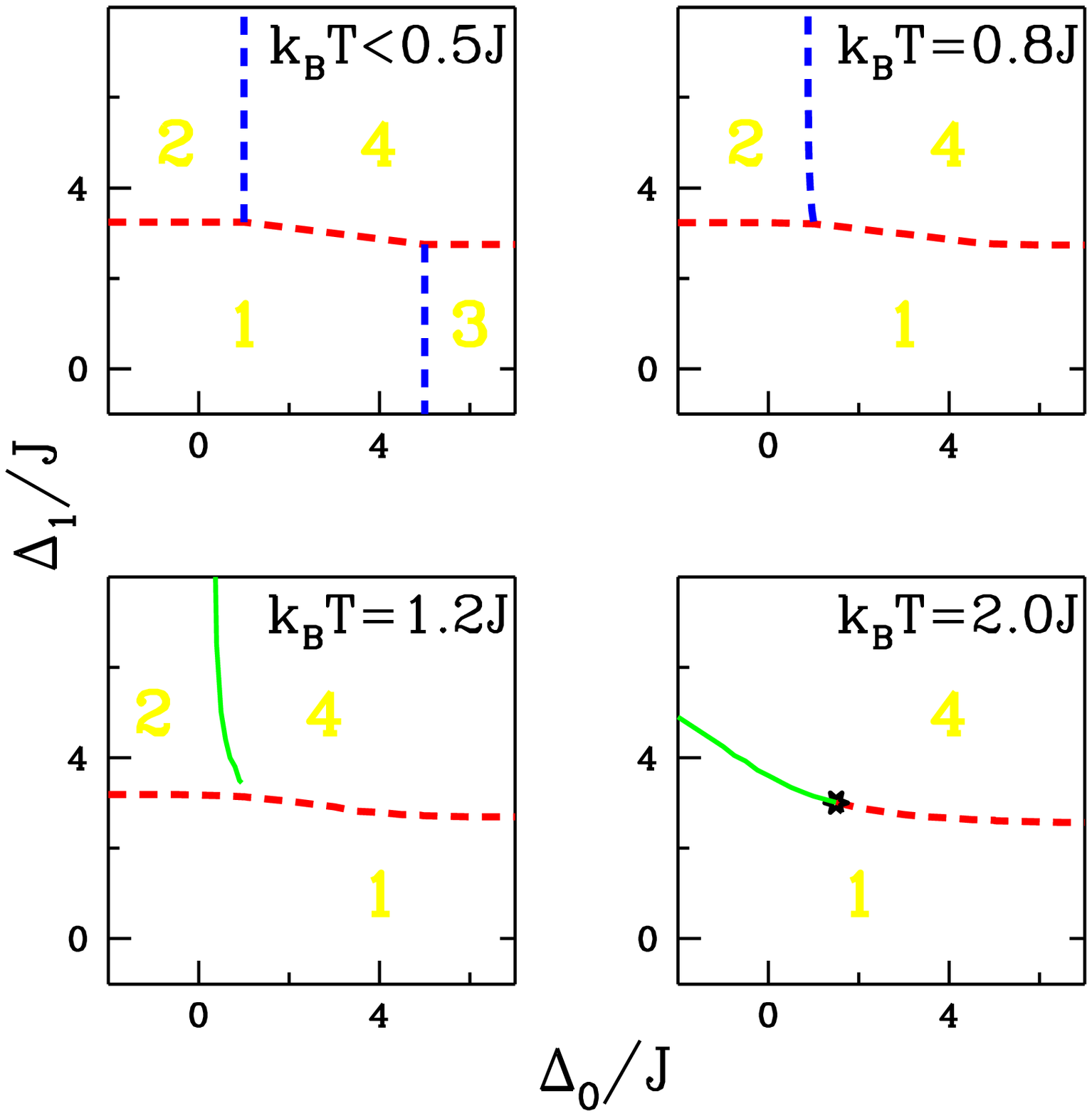}}
%\vspace*{1.8cm}
\vspace*{2.3cm}
\caption{ }
\end{figure}

%FIGURE 9
\begin{figure}
\vspace*{1.5cm}
\epsfxsize=4.5in
\hspace*{-0.5cm}
\centerline{\epsffile{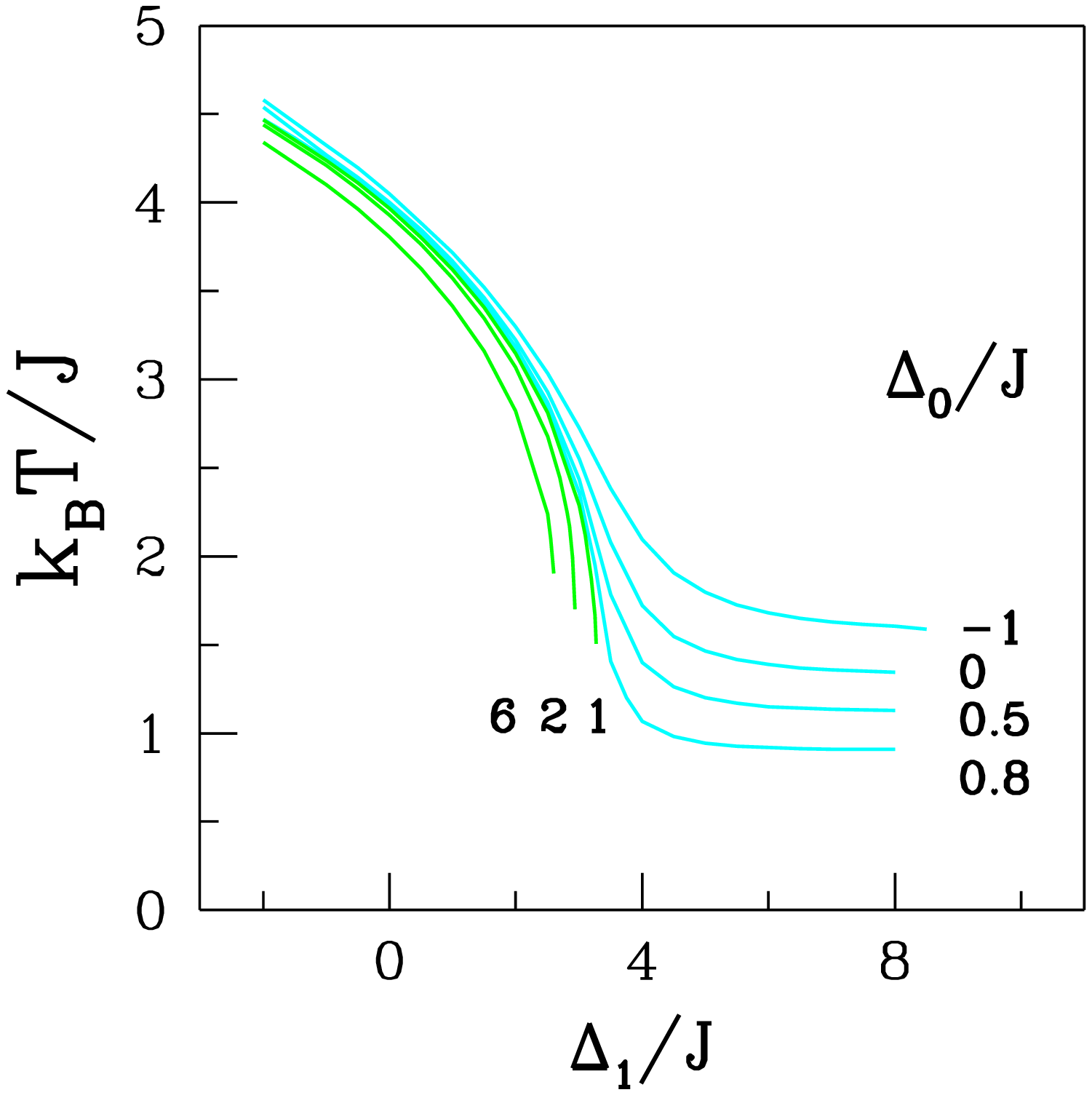}}
%\vspace*{1.8cm}
\vspace*{2.3cm}
\caption{ }
\end{figure}

%\vspace*{-3cm}
%{\bf Uniform BEG -- the diagonal cut}

%\vspace*{-1.cm}
%FIGURE 10
\begin{figure}
\vspace*{-2.0cm}
\epsfxsize=6.5in
\hspace*{-0.5cm}
\centerline{\epsffile{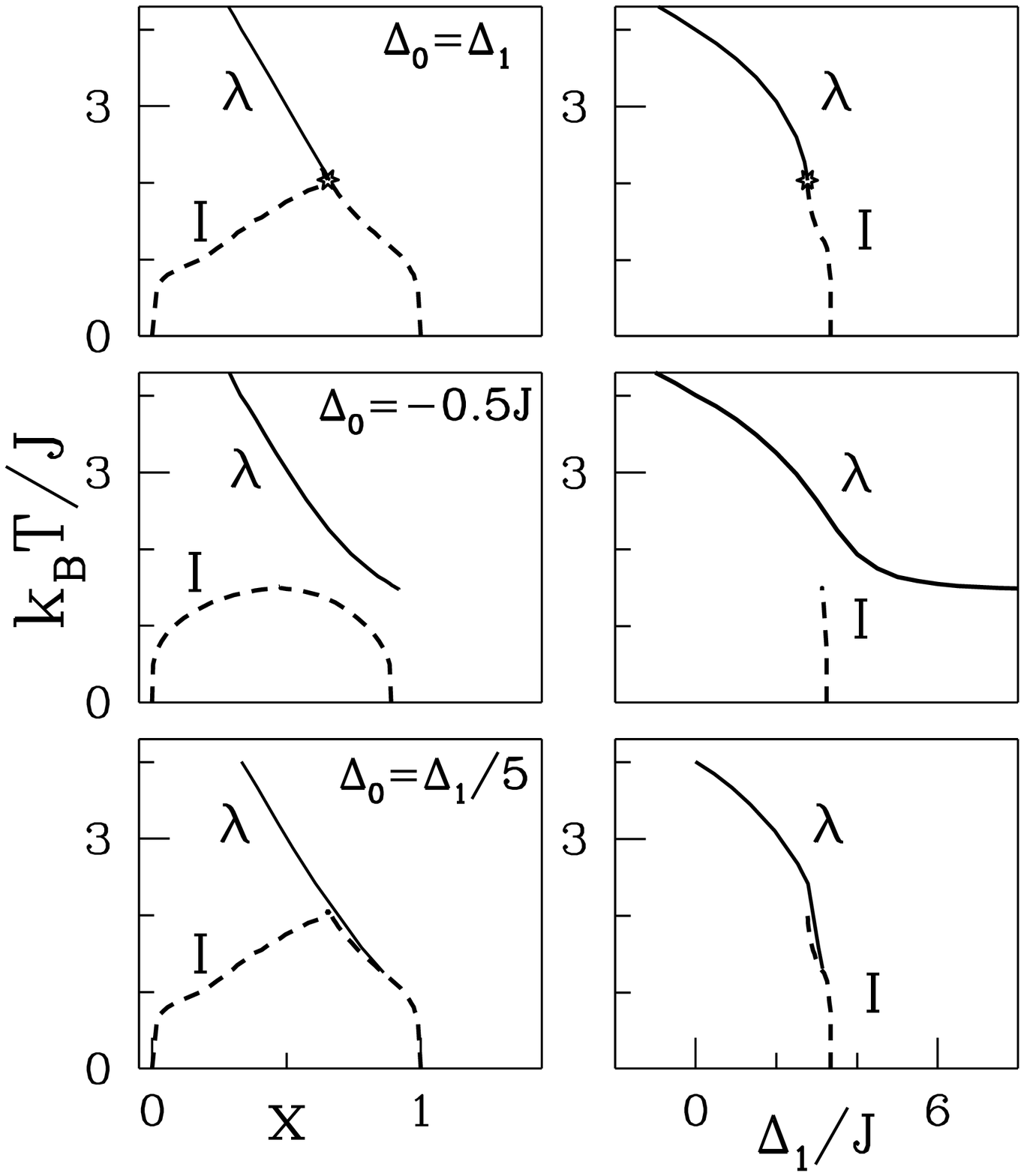}}
%\vspace*{1.8cm}
\vspace*{2.1cm}
\caption{ }
\end{figure}

%FIGURE 11
\begin{figure}
\vspace*{-2cm}
\epsfxsize=4.2in
\hspace*{-0.5cm}
\centerline{\epsffile{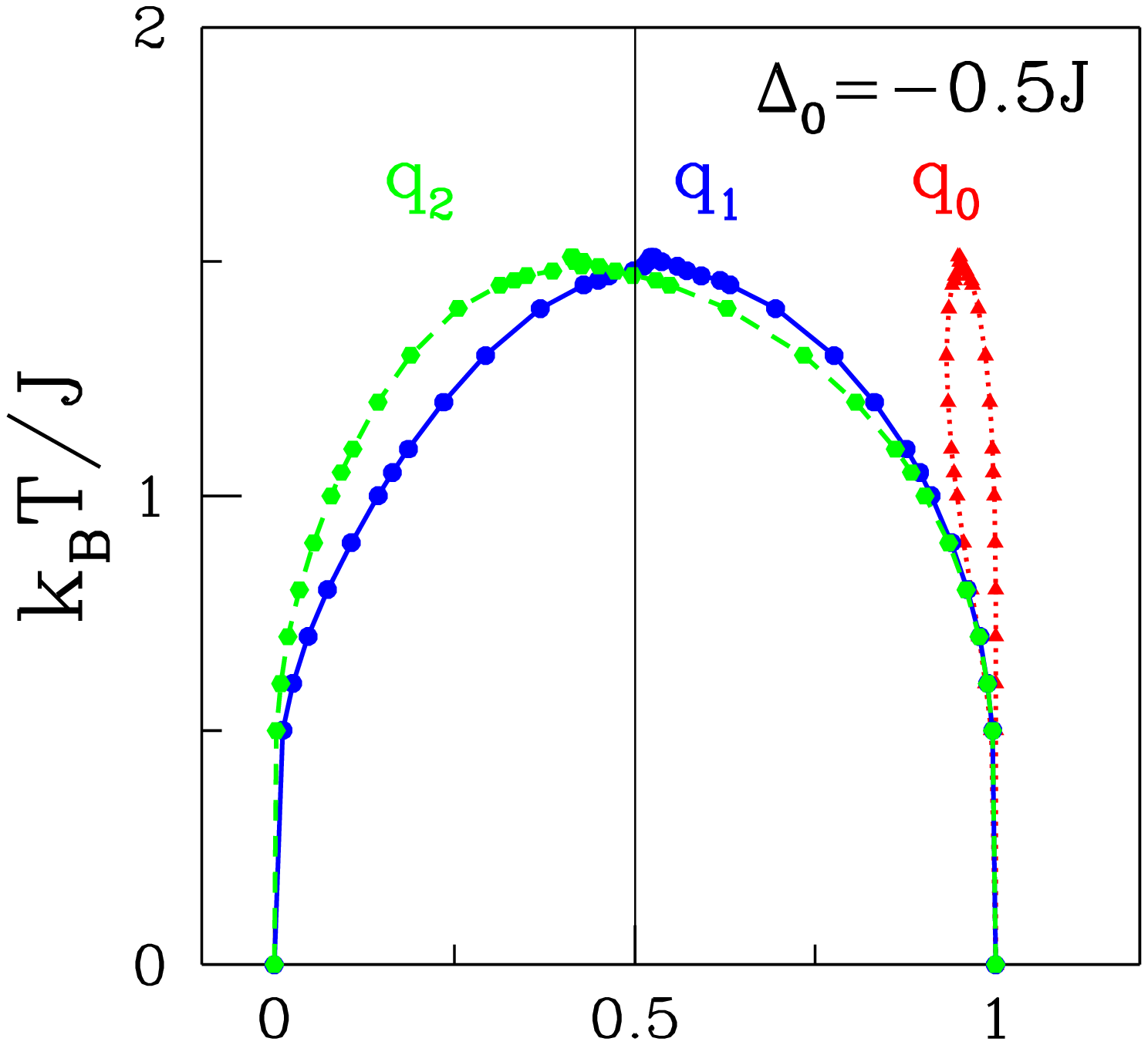}}
%\vspace*{1.8cm}
\vspace*{1.3cm}
\caption{ }
\end{figure}

\end{document}